\documentclass[10pt,conference]{IEEEtran}
\IEEEoverridecommandlockouts

\usepackage{cite}
\usepackage{subfigure}
\usepackage{amsmath,amssymb,amsfonts}
\usepackage{algorithmic}
\usepackage{graphicx}
\usepackage{textcomp}
\usepackage{xcolor}
\usepackage{float}
\usepackage[a4paper,top=0.8in,bottom=4.41cm,left=0.625in,right=0.625in]{geometry}

\def\BibTeX{{\rm B\kern-.05em{\sc i\kern-.025em b}\kern-.08em
    T\kern-.1667em\lower.7ex\hbox{E}\kern-.125emX}}
\begin{document}

\title{Cross-Layer-Optimized Link Selection for Hologram Video Streaming over Millimeter Wave Networks
\thanks{This work was supported in part by National Natural Science Foundation of China under grant No. 62371450 and the Cooperation Project Between Chongqing Municipal Undergraduate Universities and Chinese Academy of Sciences under grant HZ2021015.}
}

\author{

\IEEEauthorblockN{
Yiming Jiang\IEEEauthorrefmark{1}\IEEEauthorrefmark{2},
Yanwei Liu\IEEEauthorrefmark{1}\IEEEauthorrefmark{2}\IEEEauthorrefmark{5}, 
\thanks{\IEEEauthorrefmark{5}Corresponding Author: Yanwei Liu (Email: liuyanwei@iie.ac.cn)}
Jinxia Liu\IEEEauthorrefmark{3},
Antonios Argyriou\IEEEauthorrefmark{4},
Yifei Chen\IEEEauthorrefmark{1}\IEEEauthorrefmark{2},
and
Wen Zhang\IEEEauthorrefmark{1}}

\IEEEauthorblockA{\IEEEauthorrefmark{1}State Key Laboratory of Cyberspace Security Defense, Institute of Information Engineering, \\ Chinese Academy of Sciences, Beijing, China}
\IEEEauthorblockA{\IEEEauthorrefmark{2}School of Cyber Security, University of Chinese Academy of Sciences, Beijing, China}
\IEEEauthorblockA{\IEEEauthorrefmark{3}Zhejiang Wanli University, Ningbo, China}
\IEEEauthorblockA{\IEEEauthorrefmark{4}Department of Electrical and Computer Engineering, University of Thessaly, Greece\\}
\vspace{-1cm}

}

\maketitle

\begin{abstract}
Holographic-type communication brings an immersive tele-holography experience by delivering holographic contents to users. As the direct representation of holographic contents, hologram videos are naturally three-dimensional representation, which consist of a huge volume of data. Advanced multi-connectivity (MC) millimeter-wave (mmWave) networks are now available to transmit hologram videos by providing the necessary bandwidth. However, the existing link selection schemes in MC-based mmWave networks neglect the source content characteristics of hologram videos and the coordination among the parameters of different protocol layers in each link, leading to sub-optimal streaming performance. To address this issue, we propose a cross-layer-optimized link selection scheme for hologram video streaming over mmWave networks. This scheme optimizes link selection by jointly adjusting the video coding bitrate, the modulation and channel coding schemes (MCS), and link power allocation to minimize the end-to-end hologram distortion while guaranteeing the synchronization and quality balance between real and imaginary components of the hologram. Results show that the proposed scheme can effectively improve the hologram video streaming performance in terms of PSNR by 1.2dB to 6.4dB against the non-cross-layer scheme.
\end{abstract}


\section{Introduction}
Holography [1] is a technique that uses the principles of interference and diffraction to record the amplitude and phase information of reflected and transmitted light waves from objects, thereby enabling the reproduction of true three-dimensional images. This innovative approach allows for the creation of holograms that can be viewed from different angle, offering an immersive visual experience compared to traditional imaging methods. Thus, it can be widely used in medicine, education, and recently the metaverse.

To enable tele-holography, 
Naughton et al. [2] studied the compressibility of phase-shifted digital holograms using several lossless compression algorithms. He found that storing digital holograms in an intermediate encoding format of separate data streams of real and imaginary components can achieve better compression efficiency. This improvement in compression technique can enhance significantly the efficiency of digital hologram communication by reducing the data size while preserving image quality. Therefore, real and imaginary data are now adopted as the common data form for effective compression during hologram transmission [3].

Unlike conventional videos, digital holograms contain more information since they are three-dimensional content that records the amplitude and phase information of objects. Hence their huge data volumes require considerable bandwidth for transmission even after compression. This poses a significant challenge to modern bandwidth-limited communication networks for obtaining a perfect holography experience. As communication technology continues to advance, the demand for holographic-type communication is also increasing. This growing demand highlights the need for more effective transmission technologies which can deal with the substantial bandwidth requirements of holographic video.

With the evolution of wireless communication systems, millimeter-wave (mmWave) networking [4] stands to benefit from higher frequency spectrum resources and so it supports larger bandwidth. This capability fully aligns with the need of holographic streaming and allows us to transmit a large volume of hologram data efficiently with low latency. To overcome the issue of obstacle occlusions in mmWave communication, the multi-connectivity (MC) architecture is continuously optimized [5]. Also, the multi-connectivity channel model under mobile obstruction environments has also been extensively studied [6]. These advancements in mmWave technology led to improved transmission speed and reliability, making it a promising solution for dealing with the challenges associated with hologram streaming.

Even with MC-based mmWave networks, the transmission of hologram videos still confronts an crucial problem, namely the transmission link selection. In the past, the available link selection approaches were solely based on link quality ranking. This method neglects the adaptation of source content to links and also the harmonization among the parameters of different protocol layers in each link, something that leads to sub-optimal streaming performance. In addition, hologram videos require simultaneous transmission of real and imaginary parts. Such two channels of data give rise to the issues such as synchronization and quality balance among them. 

Considering the complex nature of a hologram image, in this paper we propose a cross-layer-optimized link selection framework for hologram video streaming over MC-based mmWave networks. It holistically considers the source coding bitrate selection in the application layer, link modulation and channel coding scheme (MCS) selection in the physical layer, and link power allocation for minimizing the end-to-end hologram distortion under the constraint of transmitting power. In the mean time, the synchronization and quality balance among real and imaginary parts in the complex-plane are fully ensured in the link selection optimization by using a distortion gap constraint.

The rest of this paper is organized as follows. Section II presents the related work. Section III provides cross-layer-optimized link selection framework for hologram video streaming. Section IV simulates the transmission process and evaluates the streaming performance. Finally, Section V concludes the paper.

\section{Related Work}
\subsection{Holographic-Type Communication}
To deliver holography experience to users, there are basically two types of hologram data representations suitable for streaming. The first delivers image-based and volumetric-based representations [7], and then generates the holograms in the client. But hologram generation has huge computational complexity that is almost not affordable by modern commonly used devices. Another type of representation directly conveys the computer-generated holograms (CGH) [8] or optically-captured holograms for a user. The hologram can be generated in advance in an off-line way. Directly delivering hologram can avoid the complicate computation needed for hologram generation at the client side, but it still demands considerable bandwidth for transmission [9]. In this work we focus on the CGH video streaming by optimizing mmWave link selection to cope with its substantial bandwidth requirements.  

\subsection{Video Streaming over MmWave Networks}
By using the high-throughput capacity of mmWave networks, several video streaming schemes have been proposed [10]. To enhance the streaming reliability, multi-connectivity architectures [11] can be leveraged for mmWave networking. To cater for the characteristics of VR video, the computation offloading scheme via mmWave channels [12] was also devised to optimize VR streaming quality of experience (QoE). As an emerging type of media, hologram video streaming over mmWave networks is still unexplored. 

\subsection{Cross-layer Optimized Video Streaming}
Several cross-layer optimization methods [13] were proposed to ensure the source-channel rate adaptation for video streaming. By jointly selecting the parameters in the application layer, MAC layer and physical layer towards end-to-end QoE optimization [14], the video streaming performance can be significantly improved compared to the non-cross-layer scheme. When cross-layer optimization takes place to the holographic video streaming over MC-based mmWave networks, the presence of two channels for the real and imaginary components in complex plane requires  the investigation of dynamic selection of mmWave links for improved quality. 

\section{Cross-Layer-Optimized Link Selection Framework}

\subsection{Framework Overview}
Fig.1 illustrates the proposed cross-layer-optimized mmWave link selection framework. The system consists of an optimization controller which resides in a sub-6 GHz gNB (next generation Node B) that can directly communicate with several slave mmWave gNBs, different gNBs that allocate the corresponding physical layer MCSs to the client and a hologram content server that contains multiple copies of the hologram video segments with various bit-rates (encoded with different quantization parameters (QPs)). The optimization controller is expected to minimize the end-to-end distortion for the transmitted hologram segment over the optimally selected links. To this end, the link MCS, link power allocation and the hologram coding QP are jointly optimized by the controller in a holistic manner.

\begin{figure}[htbp]
\vspace{-0.2cm}
\centerline{\includegraphics[width=3in]{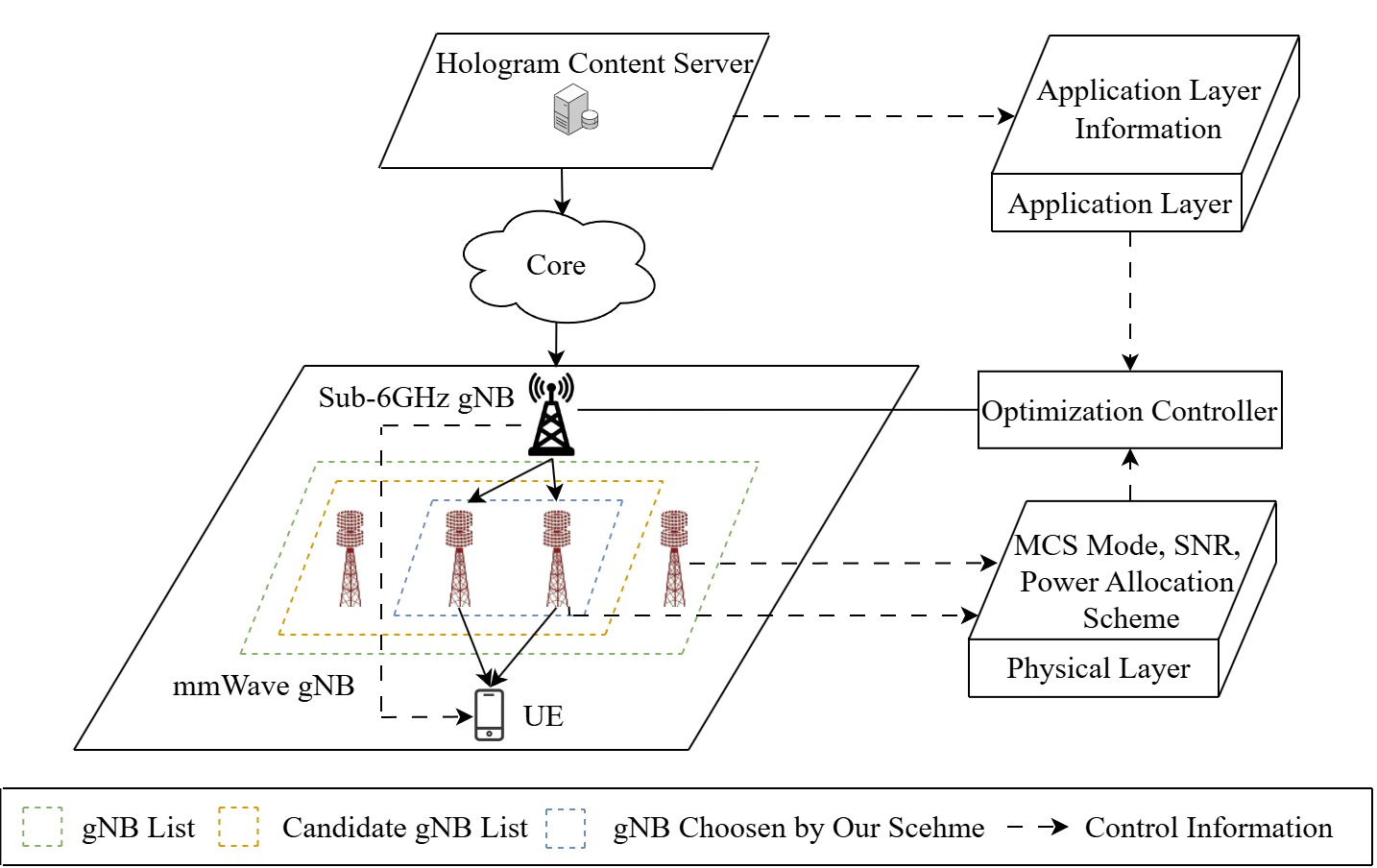}}
\vspace{-0.2cm}
\caption{Cross-layer-optimized mmWave link selection framework.}
\label{fig_1}
\end{figure}

\vspace{-0.3cm}

\subsection{Communication Model}
\subsubsection{MC-based mmWave communication} Due to the natural data partitioning features of real and imaginary parts of hologram video and the unsteady characteristics of mmWave links, it is better to select at least two suitable links for transmitting large volumes of real and imaginary data simultaneously for a UE. To achieve this, the optimization controller should consider not only the establishment of multiple available mmWave links but also the optimization of dynamic network parameters for each selected link. Thus, the hologram video steaming framework can adapt to the link with the appropriately periodic updating of links.

Usually, there might be multiple mmWave links connecting to a multi-homed UE, thus requiring a basic assessment of link availability before selection. Since the mmWave networks tend to be noise-limited rather than interference-limited [15] due to highly directional beamforming and sensitivity to blockages, the instantaneous signal-to-noise-ratio (SNR) value from gNB to UE is used for preliminary link screening. The optimization controller that is located at the sub-6 GHz gNB takes charge of scheduling the transmission time slots to different links. The UE broadcasts the uplink sounding reference signals to the mmWave gNB. The uplink sounding reference signal is then used to estimate the mmWave downlink quality. The mmWave gNB aligns its beam direction with the UE to find the best transmission direction, and next sends the corresponding SNR value to the controller. The controller reports the optimal transmission direction and its mmWave link SNR value to the UE via a reliable sub-6GHz link. 

We use a minimum SNR threshold value ${S_T}$ to build the candidate mmWave gNB list $L$ from the entire gNB list for a UE. When the SNR value of link $i$ is greater than ${S_T}$, it is considered that this link can be used as a candidate link for hologram video transmission. In mmWave transmission, power allocation between the links significantly affects the channel SNR and finally affects the received hologram video quality. Therefore, specific power allocation among links within the constraint of total transmitting power can be considered as an optimizable parameter for hologram video streaming.

\subsubsection{Blockage model}
Since the mmWave link is prone to be blocked by human activities, we introduce the human blockage model [16] to the MC-based mmWave communication. The model simulates potential moving blockage event for each link by using a Poison Point Process with dynamic blocker density $\lambda_B$  within a certain circular region. The specific schematic diagram for human blocking is shown in Fig. 2.When the blocker reaches the position near a gNB-UE link, the blockage event will cause transmission power loss, and will last for a certain period. Hence, each time a blockage event occurs, the received power drops by the average human blockage loss [16], which is defined by the blockage event time duration. The blockage power loss further affects the link SNR. Then the transmission recovers to normal until the blockage moves away. 
\begin{figure}[htbp]
\vspace{-0.3cm}
\centerline{\includegraphics[width=1.4in]{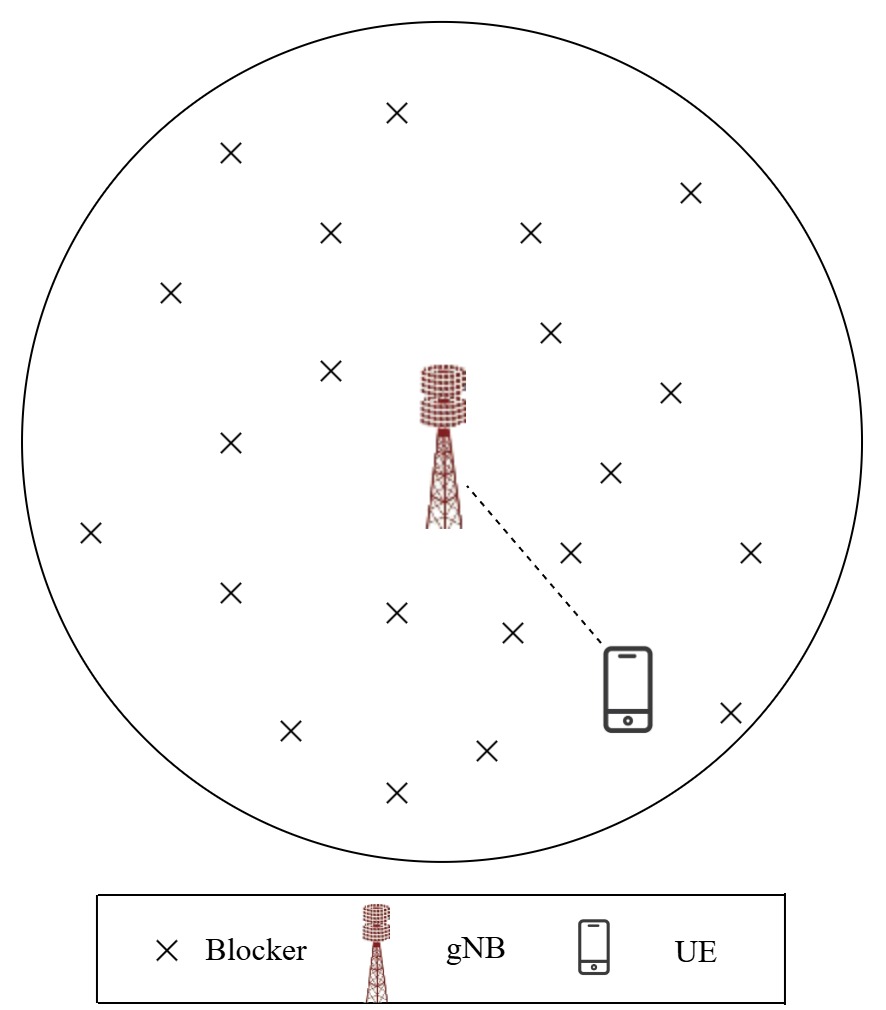}}
\vspace{-0.2cm}
\caption{Human blockage diagram.}
\label{fig_1}
\vspace{-0.2cm}
\end{figure}
\subsubsection{MCS adaptation of the mmWave link} At the physical layer of mmWave links, different MCS can be selected to provide different error-resilience and transmission capacities. In a mmWave transmission slot, the transmission block (TB) consists of several coding blocks (CBs).Accordingly, the coding block error rate (BLER) directly affects the received video quality by the UE. Based on the mean mutual information per coded bit (MMIB) for each CB, the BLER for each CB can be formulated with a Gaussian cumulative model [14],
\begin{equation}
\label{eqn_1}
{C_{B,i}}({\beta _i}) = \frac{1}{2}\left[ {1 - erfc(\frac{{{\beta _i} - b(m)}}{{\sqrt 2 c(m)}})} \right]
\end{equation}
where ${C_{B,i}}({\beta _i})$ and ${\beta _i}$ are the BLER and MMIB for the ${i^{th}}$ block in the TB, respectively. The parameters $b(m)$ and $c(m)$ are the mean and standard deviation of Gaussian distribution [17], and $m$ denotes the selected MCS mode. Then, the BLER for one TB with ${N_T}$ CBs is 
\begin{equation}
\label{eqn_2}
{T_B} = 1 - \prod\nolimits_{i = 1}^{{N_T}} {(1 - {C_{B,i}}({\beta _i}))} 
\end{equation}
Next, the hologram packet loss rate can thus be predicted as:
\begin{equation}
\label{eqn_3}
{p_{n,i}} = 1 - \prod\limits_{i = 1}^{{N_B}} {(1 - {T_{B,i}}))} 
\end{equation}
where ${N_B}$ is the number of TBs that the hologram data packet contains,  and ${T_{B,i}}$ denotes the BLER of the ${i^{th}}$ TB. 

\subsection{Distortion Model}\label{AAA}
Assume that the hologram content server contains $Q$ quality versions of hologram video with different coding QPs. Each quality version consists of a series of segments. To perform the cross-layer-optimized link selection, we need to estimate the end-to-end hologram distortion under different parameter. Considering the practical video packaging process, video segment is partitioned into Groups of Pictures (GOPs) during compression, with each GOP containing multiple packets. Since error control among packets within a GOP introduces correlations that affect each packet, estimating the distortion of a GOP requires integrating the distortions of its individual packets. 


Mean Squared Error (MSE) measures the average difference between the distorted and true images, which is used to assess the video distortion. For a GOP with the vector of packet loss rate being $\mathcal{\gamma}  = \{ {p_0},...,{p_i},...{p_{N-1}}\}$, its end-to-end distortion in term of MSE can be derived as
\begin{equation}
\label{eqn_4}
\begin{array}{l}
{D_\mathcal{\gamma}}(GOP) = \sum\limits_{k = 0}^{{2^N} - 1} {{p^{(k)}}{D^{(k)}}} \\
\;\;\;\;\;\;\;\;\;\;\;\;\;\;\; \;\;  = \sum\limits_{k = 0}^{{2^N} - 1} {(\prod\limits_{i = 0}^{N - 1} {{{(1 - {p_i})}^{(1 - b_i^{(k)})}}p_i^{b_i^{(k)}}} )} {D^{(k)}}
\end{array}
\end{equation}
where $D^{(k)}$ denotes the distortion of a whole GOP when $p^{(k)}$ occurs and $k=0,1,...,2^N-1$. $p^{(k)}$ is the probability that the $k^{th}$ event vector $\mathcal{B}_{(k)}=\{b_0^{(k)},...b_i^{(k)},...b_{N-1}^{(k)}\}$ occurs, where $b_i=0$ to indicate the $i^{th}$ packet has been received unsuccessfully and $b_i=1$ to indicate the $i^{th}$ packet has been received successfully. $N$ indicates the number of packets in the GOP. $p_i$ is the packet loss probability of the $i^{th}$ packet in the GOP, that can be estimated by Eq. (3). 

Using the approach above may increase the computational complexity, which is inappropriate for timely computation of distortion that is crucial for hologram streaming. 
To address this issue, we opted to estimate the first order distortion [18] using Taylor expansion to simplify the actual computational process. For a GOP with the packet loss rate vector $\mathcal{\gamma}$, its end-to-end distortion can be approximated by Taylor expansion as:
\begin{equation}
\label{eqn_5}
\begin{array}{l}
{D_\mathcal{\gamma}}(GOP) \approx  {D_{\mathcal{\bar\gamma}}}(GOP) + \sum\limits_{i = 0}^{N - 1} {\frac{{\partial {D_\mathcal{\gamma}}(GOP)}}{{\partial {p_i}}}} {|_{\mathcal{\gamma} =\mathcal{\bar\gamma}}}({p_i} - {{\bar p}_i})\\
\;\;\;\;\;\;\;\;\;\;\;\;\;\;\;\;\;= {D_{\mathcal{\bar\gamma}}}(GOP) + \sum\limits_{i = 0}^{N - 1} {{\lambda _i}} ({p_i} - {{\bar p}_i})
\end{array}
\end{equation}
where $\mathcal{\bar\gamma} = \{0,0,...,0\} $ denotes the packet loss rate vector where Taylor expansion occurs, and $p_i$, $\bar p_i$ denote the effective packet loss rate and the reference
packet loss rate of the $i^{th}$ packet. According to the law of total probability, $D_\mathcal{\gamma}(GOP)$ can be transformed into the following form,
\begin{equation}
\label{eqn_6}
{D_\mathcal{\gamma}}(GOP) = (1 - {p_i}){D_\mathcal{\gamma}}(GOP){|_{{b_i} = 1}} + {p_i}{D_\mathcal{\gamma}}(GOP){|_{{b_i} = 0}}
\end{equation}
Based on Eq.(6), ${\lambda _i}$ can be simplified as 
\begin{equation}
\label{eqn_7}
{\lambda _i} = {D_\mathcal{\gamma}}(GOP){|_{{b_i} = 0}} - {D_\mathcal{\gamma}}(GOP){|_{{b_i} = 1}}
\end{equation}
where ${D_\mathcal{\gamma}}(GOP){|_{{b_i} = 1}} = {D_{\mathcal{\bar\gamma}}}(GOP)$ when $\mathcal{\bar\gamma} = \{ 0,0,...,0\} $. And $D_\mathcal{\bar\gamma}(GOP)$ is the source encoding distortion per GOP, which can be obtained easily during hologram encoding in advance. 
Hence, by making Taylor expansion at $\mathcal{\bar\gamma}$, we can estimate the distortion of the real and imaginary parts for hologram video streaming over mmWave links.

\subsection{Link Selection with Cross-layer Optimization}\label{ITH}
For the MC-mmWave communication, the path loss and human blockage affect significantly the state of transmission links. Dynamic link selection with the adjustments of network parameters of difficult protocol layers is very necessary to ensure the performance of hologram streaming. When performing link selection optimization, the first objective is maximizing the received hologram quality that is implemented by minimizing the end-to-end hologram distortion ${D_H}$ by configuring optimal network parameters of different protocol layers. While reducing the distortion of hologram video, the objective for controlling synchronization and quality balance between real and imaginary parts should also be taken into account. Only well-balanced quality between real and imaginary parts can result in good hologram reconstruction quality. Hence, the cross-layer optimization problem is derived as  Eq.(8), where $l_R$ and $l_I$ denote the selected links from candidate link list $L$ for real and imaginary parts, respectively. $q_R$ and $q_I$ are the selected QPs for the real and imaginary parts over the  $l_R$ and $l_I$ links, respectively. $m_R$ and $m_I$ are the MCS modes selected from the candidate set $M$ for $l_R$ and $l_I$ links, respectively. ${\Omega _R}$ and ${\Omega _I}$ are the powers allocated to $l_R$ and $l_I$ links, and ${\Omega}$ is the overall transmitting power used for hologram video streaming. ${D_H}$ denotes the hologram distortion that includes the real part distortion ${D_R}$ and imaginary part distortion ${D_I}$. ${D_T}$ is the maximum threshold value that can maintain synchronization and quality balance between the real and imaginary parts of the hologram.
\begin{equation}
\label{eqn_8}
\begin{array}{l}
q_R^{opt},m_R^{opt},\Omega _R^{opt},l_R^{opt}, q_I^{opt},m_I^{opt},\Omega _I^{opt},l_I^{opt}\\
 = \mathop {\arg \min }\limits_{\scriptstyle{q_{R}},\;{q_I} \in \;Q\hfill\atop
{\scriptstyle{m_R},{m_I} \in M\hfill\atop
\scriptstyle{l_R},{l_I} \in L\hfill}} \;{D_H}({q_R},{m_R},{\Omega _R},l_R^{},{q_I},{m_I},{\Omega _I},l_I^{})\\
 = \mathop {\mathop {\arg \min }\limits_{\scriptstyle{q_{R}},\;{q_I} \in \;Q\hfill\atop
{\scriptstyle{m_R},{m_I} \in M\hfill\atop
\scriptstyle{l_R},{l_I} \in L\hfill}} }\limits_{} \;({D_R}({q_R},{m_R},{\Omega _R},l_R^{}) + {D_I}({q_I},{m_I},{\Omega _I},l_I^{}))\;\;\;\;\;\;\;\;\;\;\;\;\\
{\rm{s}}{\rm{.t}}{\rm{.  }}\left| {{D_R}({q_R},{m_R},{\Omega _R},l_R^{}) - {D_I}({q_I},{m_I},{\Omega _I},l_I^{})} \right| \le {D_T},\\
\;\;\;\;\;\;{\Omega _R} + {\Omega _I} \le \Omega 
\end{array}
\end{equation}

Based on the optimization formulation Eq.(8), we can obtain the optimal links and their corresponding network parameters by iteratively searching through the different parameter combinations. 

\begin{figure*}[ht]
\vspace{-0.3cm}
\centering
\subfigure[${\lambda _B} = 0.03bl/{m^2}$]{
\label{fig:subfig:a}
\includegraphics[width=2in]{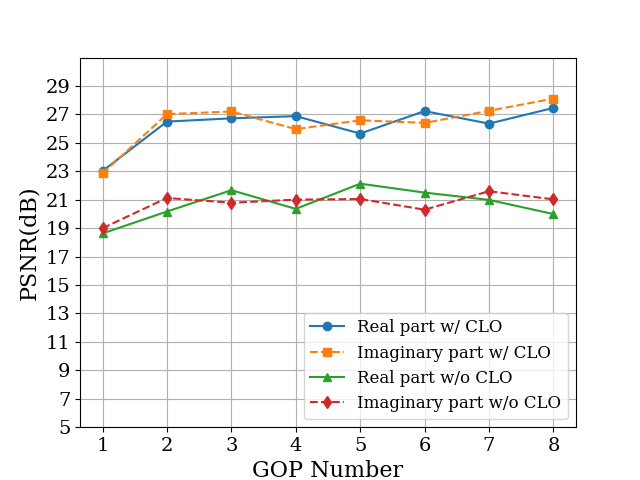}}
\subfigure[${\lambda _B} = 0.05bl/{m^2}$]{
\label{fig:subfig:b} 
\includegraphics[width=2in]{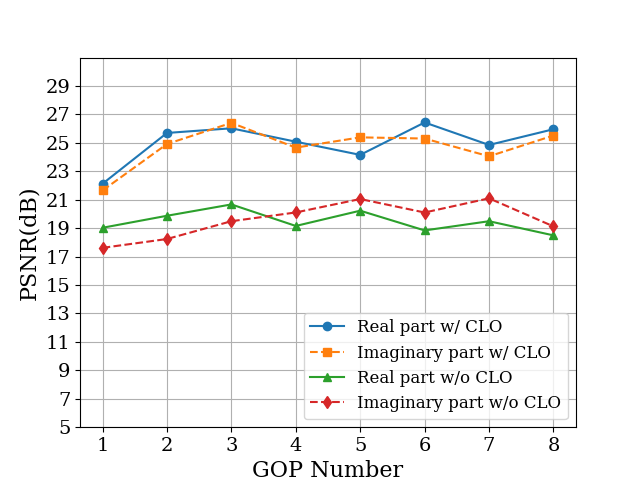}}
\subfigure[${\lambda _B} = 0.1bl/{m^2}$]{
\label{fig:subfig:c} 
\includegraphics[width=2in]{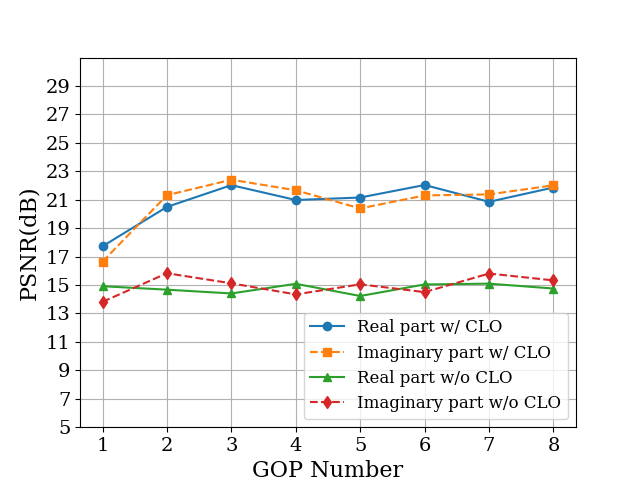}}
\vspace{-0.2cm}
\caption{Hologram video (BreakDancer) streaming performance in terms of PSNR under different blocker densities.}
\label{fig:subfig} 
\end{figure*}
\begin{figure*}[ht]
\vspace{-0.5cm}
\centering
\subfigure[${\lambda _B} = 0.03bl/{m^2}$]{
\label{fig:subfig:a}
\includegraphics[width=2in]{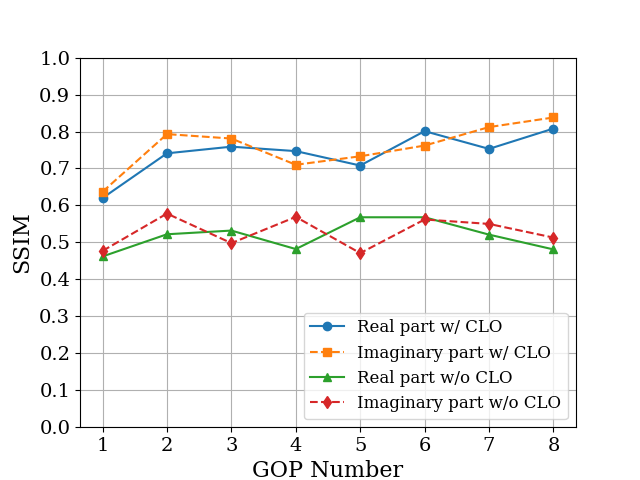}}
\subfigure[${\lambda _B} = 0.05bl/{m^2}$]{
\label{fig:subfig:b} 
\includegraphics[width=2in]{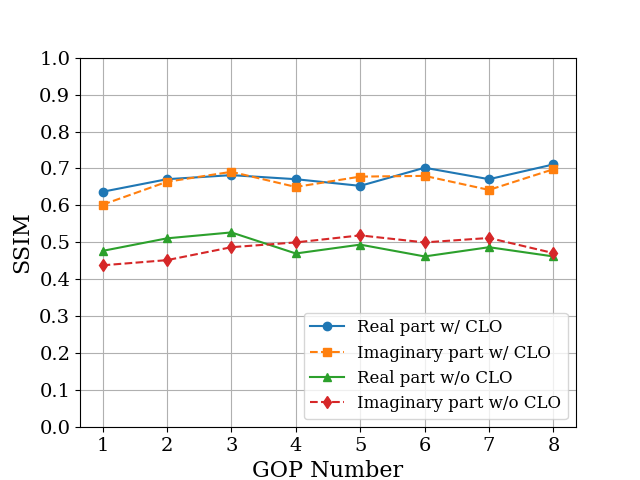}}
\subfigure[${\lambda _B} = 0.1bl/{m^2}$]{
\label{fig:subfig:c} 
\includegraphics[width=2in]{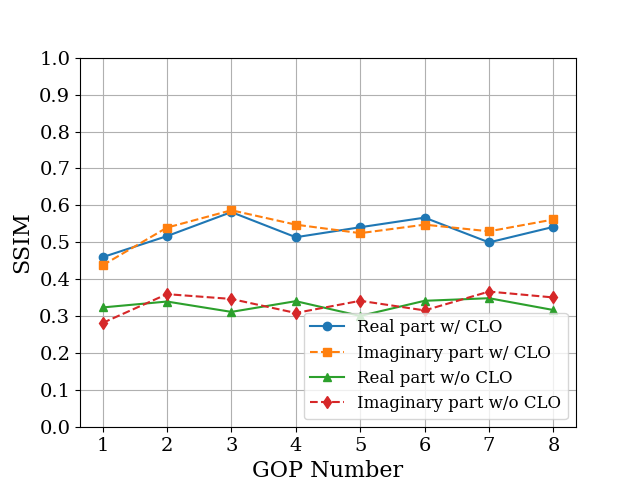}}
\vspace{-0.2cm}
\caption{Hologram video (BreakDancer) streaming performance in terms of SSIM under different blocker densities.}
\label{fig:subfig} 
\vspace{-0.6cm}
\end{figure*}
\vspace{-0.3cm}
\begin{table}[htbp]
\caption{SIMULATION PARAMETERS}
\vspace{-0.4cm}
\begin{center}
\begin{tabular}{|c|c|}
\hline
Coding structure & IPPP \\
\hline
Candidate QP & 27,37,45 \\
\hline
Distance between node (m) & 10 \\
\hline
Number of mmWave links & 6 \\
\hline
frequency & 30GHz  \\
\hline
Total txPower  & 60dBm\\
\hline
Noise figure (dB)   & 9dB \\
\hline
mmWave simulation Scenario  & UMi \\
\hline
Height of UE  & 1.5m \\
\hline
Height of gNB & 10m  \\
\hline
{SNR threshold $S_T$}   &  15dB \\
\hline
Human blockage density (bl/$m^2$) & 0.03, 0.05, 0.1 \\
\hline
{Distortion threshold  $D_T$} in terms of PSNR  &  1.5dB \\
\hline
\end{tabular}
\label{tab1}
\end{center}
\vspace{-0.65cm}
\end{table}

\section{Experimental Results}\label{FAT}
In our experiments, the H.265/HEVC [19] was used for encoding multiple quality versions of hologram videos. The NYUSIM [20] mmWave channel model is used to establiash multiple links to simulate the experimental environment. We use the Urban Microcell (UMi) scenario for hologram video streaming experiments since it provides large wireless coverage and network capacity to meet the demand for high-speed data transmission required by hologram video. The specific simulation parameters are shown in Table I.

To test the performance of the proposed cross-layer optimized(CLO) link selection scheme, we use BreakDancer [21] sequence with a resolution of 1920$\times$1080 and Ballet sequence with a resolution of 3840$\times$2160 for experiments. Three different QPs were used to encode the real and imaginary video streams to build various quality versions for hologram video bitrate adaptation. To enable interactive random access, each hologram segment is set to one GOP of 4 frames.

During the initial stage of our experiment, we transmit medium-quality hologram video streams, and then utilize our optimization scheme to dynamically select the links and the optimal parameters of different protocol layers. Fig. 3 and Fig. 4 show the hologram video (BreakDancer) streaming performance in terms of Peak Signal-to-Noise Ratio (PSNR)  and Structural Similarity Index Measure (SSIM) for our CLO link selection scheme against the one without cross-layer optimization (w/o CLO). For the non-CLO scheme, the medium-quality video (QP=37) is used for streaming and the links are selected by the ranking of their SNR values. To obtain reliable results, we average the results over multiple runs of experiments with each GOP under different $\lambda_B$.

In Fig. 3, the PSNR value is used to provide a more intuitive representation of streaming performance. It can be observed from Fig. 3. The hologram video quality is gradually degraded with the increasing $\lambda_B$, but with our CLO scheme, the received video quality is always higher than that without CLO for different blockage density $\lambda_B$. For the same GOP, using the CLO scheme can achieve an improvement in PSNR approximately 2.3dB to 6.4dB. The improvement shows that the proposed link selection scheme can dynamically change the links and their network parameters to accommodating the dynamics of the human blockages. In Fig.4, it can also be seen that the average SSIM value of the real and imaginary streams improved by at least 9.7\% using the proposed CLO scheme against that without CLO. This indicates that our CLO scheme can still achieve improvements in subjective quality.

Besides the quality improvement, synchronization between the real and imaginary parts during the transmission is also maintained fairly well. Specifically in Fig. 3, after applying the scheme, the difference in PSNR value between the real and imaginary parts is always less than 1dB, corresponding to that in SSIM value less than 0.05 in Fig. 4. We also note that for the scenario without CLO, the difference between real and imaginary parts is not very large too. This is because, with fewer available links, we have to select two links with good SNR to ensure the normal transmission. With more available links, the synchronization between the real and imaginary parts may be worse without CLO.

\begin{figure}[ht]
\vspace{-0.4cm}
\centering
\subfigure[\scriptsize Average PSNR]{
\label{fig:subfig:a}
\includegraphics[width=1.53in]{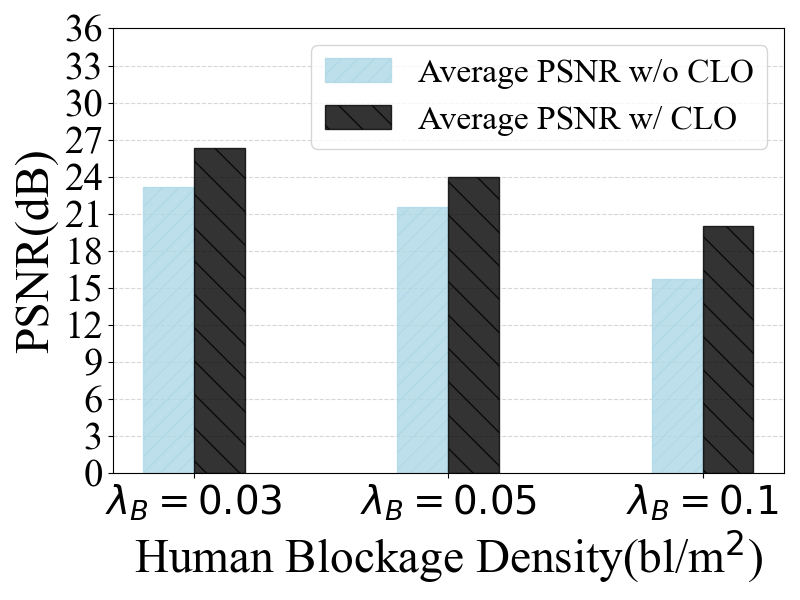} }
\subfigure[\scriptsize Average SSIM]{
\label{fig:subfig:b} 
\includegraphics[width=1.51in]{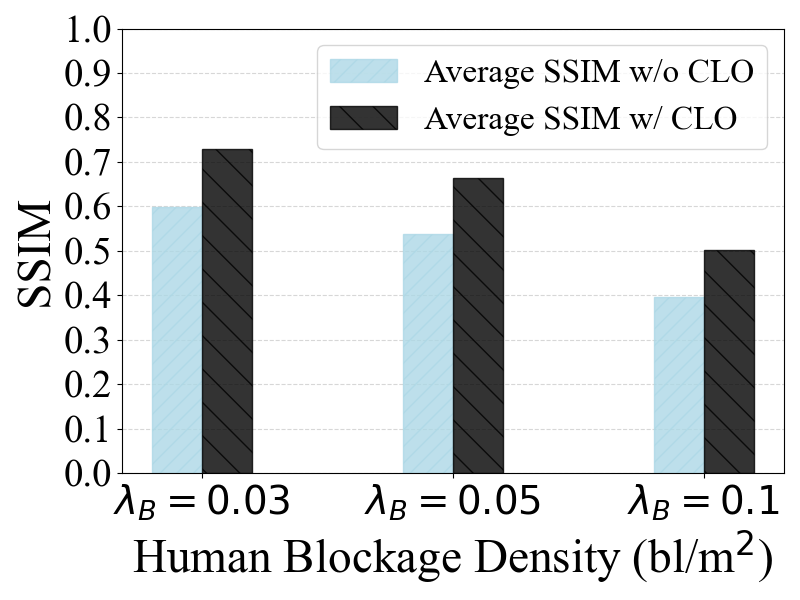}}
\vspace{-0.2cm}
\caption{Hologram video streaming performance for Ballet.}
\label{fig:subfig} 
\vspace{-0.4cm}
\end{figure}
We also conduct the experiments on large size hologram video Ballet(3840$\times$2160). The results are shown in Fig. 5. Due to the packet loss, the hologram video quality is relatively low. With the CLO scheme, the average PSNR achieves an improvement ranging from 1.2 dB to 4.1 dB. Also, the average SSIM value increases by at least 10.5\%. The results indicate that the scheme can achieve streaming quality improvements in large size hologram video.

Currently, optical hologram displays can be simulated by numerical reconstruction of the hologram that is directly visible to the viewer via a commonly used display. To further evaluate the performance of the proposed optimization scheme, we perform hologram video reconstruction by using the NRSH [7] software, allowing us to subjectively assess the visual quality of finally displayed hologram video. The specific reconstruction sample images for BreakDancer hologram are shown in Fig. 6. The amplitude images in the experiments are reconstructed with the Angular Spectrum Method at a distance of $z=0.001$m. By comparing two different amplitude images with and without the CLO scheme, it can be seen that the image without CLO is relatively blurred compared to the one with CLO. Driven by the CLO, more high frequency parts distributed all over the hologram are kept since the quality version with a higher bitrate is selected. The outlines of the dancers are clearer after CLO. This situation is more pronounced at the focus of the image. These observations illustrate that the proposed scheme effectively improves the subjective viewing quality of reconstructed hologram video.
\begin{figure}[ht]
\vspace{-0.3cm}
\centering
\subfigure[w/ CLO]{
\label{fig:subfig:a}
\includegraphics[width=2.6in]{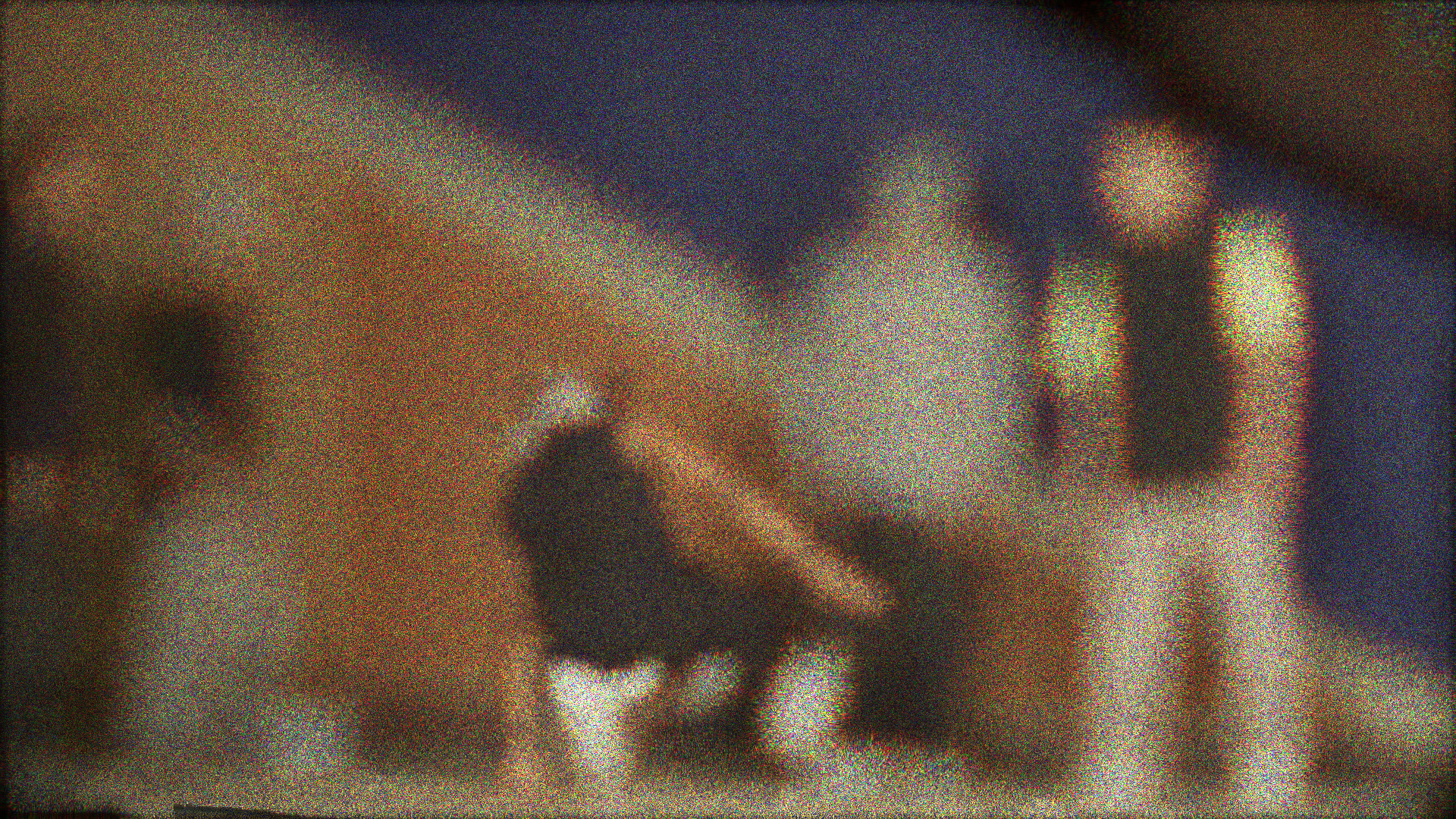}}
\subfigure[w/o CLO]{
\label{fig:subfig:b} 
\includegraphics[width=2.6in]{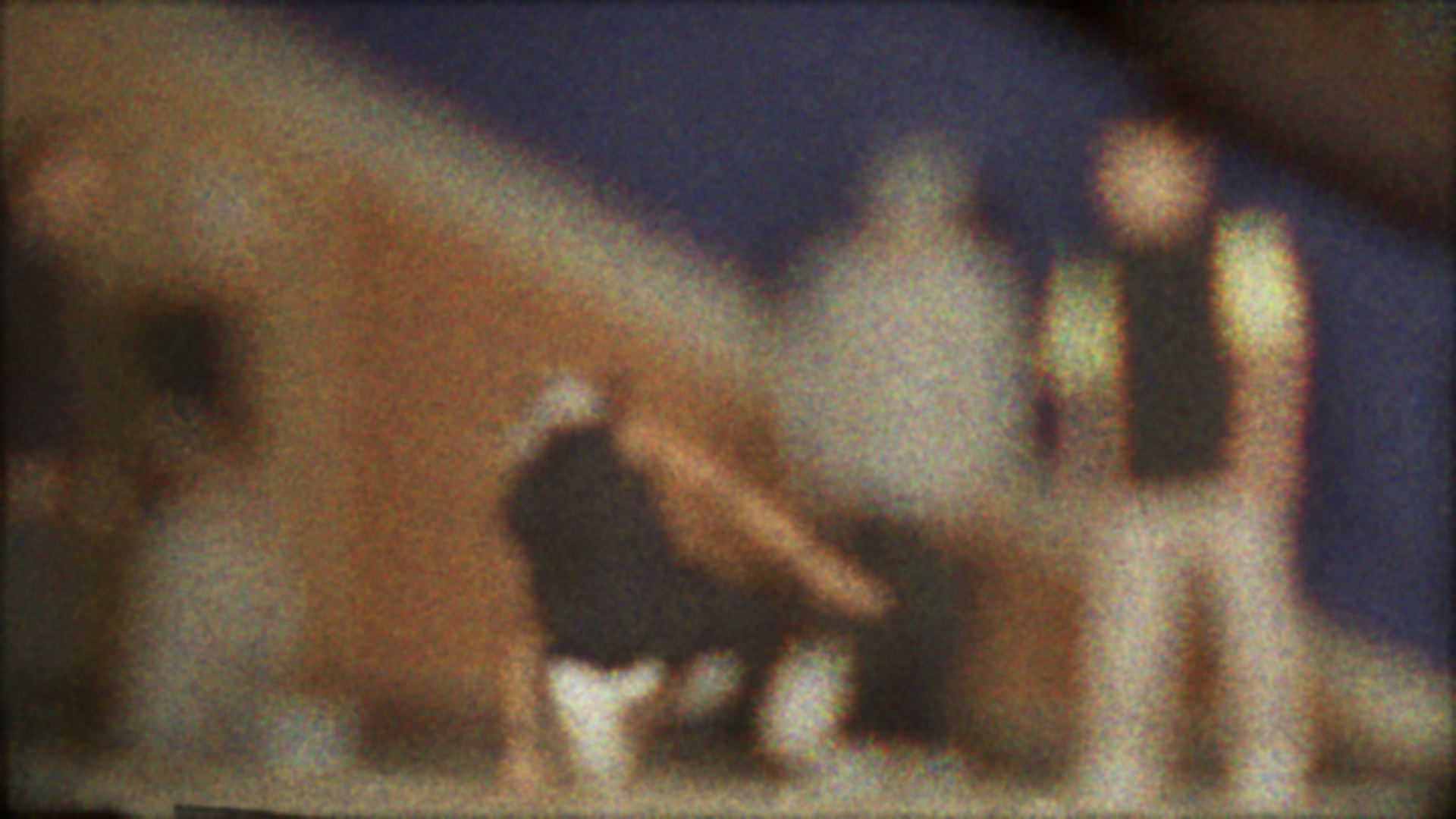}}
\caption{The comparison of reconstructed BreakDancer hologram image with and without CLO.}
\label{fig:subfig} 
\vspace{-0.6cm}
\end{figure}

\section{Conclusion}
In this paper, a cross-layer optimized link selection scheme for hologram video streaming over mmWave networks is proposed. This scheme selects links by jointly optimizing hologram coding bitrate, MCS modes, and channel power allocation to minimize distortion during the hologram video transmission while ensuring the synchronization between real and imaginary parts. We evaluated the optimization scheme by measuring the PSNR and SSIM values for two test video streams under different human blockage densities. Extensive experimental results show that the proposed scheme can effectively improve the received quality of hologram video and ensures synchronization between real and imaginary parts compared to the non-cross-layer scheme.

\vspace{12pt}

\end{document}